\documentclass[prc,onecolumn,floatfix,superscriptaddress,showpacs,nofootinbib,preprint]{revtex4}  
\usepackage{graphicx}
\usepackage{epsfig}  
\usepackage{dcolumn}
\usepackage{bm}

\begin{document}  
  
\title{Forward-backward correlations in nucleus-nucleus collisions:\\ baseline  
  contributions from geometrical fluctuations }  
\author{V.P. Konchakovski}  
\affiliation{Helmholtz Research School, University of Frankfurt, Frankfurt, Germany}  
\affiliation{Bogolyubov Institute for Theoretical Physics, Kiev, Ukraine}  
\author{M. Hauer}  
\affiliation{Helmholtz Research School, University of Frankfurt, Frankfurt, Germany}  
\author{G Torrieri}  
\affiliation{Institut f\"ur Theoretische Physik, Goethe Universit\"at,  
  Frankfurt am Main, Germany}  
\author{M.I. Gorenstein}  
\affiliation{Bogolyubov Institute for Theoretical Physics, Kiev, Ukraine}  
\affiliation{Frankfurt Institute for Advanced Studies, Frankfurt, Germany}  
\author{E.L. Bratkovskaya}  
\affiliation{Frankfurt Institute for Advanced Studies, Frankfurt, Germany}  
  
\begin{abstract}  
We discuss the effects of initial collision geometry and  
centrality bin definition on correlation and fluctuation  
observables in nucleus-nucleus collisions. We focus on the  
forward-backward correlation coefficient recently measured by the  
STAR Collaboration in Au+Au collisions at RHIC. Our study is  
carried out within two models: the Glauber Monte Carlo code with a  
`toy' wounded nucleon model and the hadron-string dynamics (HSD)  
transport approach. We show that strong correlations can arise due  
to averaging over events in one centrality bin. We, furthermore,  
argue that a study of the dependence of correlations on the  
centrality bin definition as well as the bin size may distinguish  
between these `trivial' correlations and correlations arising from  
`new physics'.  
  
\end{abstract}  
  
\maketitle  
  
\section{Introduction}  
  
Correlations of particles between different regions of rapidity  
have for a long time been considered to be a signature of new  
physics.  A shortening in the correlation length in rapidity has  
been thought to signal a transition to a quark-gluon plasma \cite{qgp1,qgp2}.  
Conversely, the appearance of long-range correlations has  been  
associated with the onset of the percolation limit, also linked to  
the QCD phase transition \cite{perc1,perc2}. Recently, the  
correlations across a large distance in rapidity have also been  
suggested to arise from a color glass condensate  
\cite{correlation_cgc1,correlation_cgc2}.   The observation of  
such  correlations in A+A collisions at RHIC energies by the  
STAR Collaboration \cite{correlation_exp1,correlation_exp2} has  
therefore elicited a lot of theoretical interest.  
  
The purpose of this work is to identify some {\em baseline}  
contributions to the experimentally observed correlations,  
contributions that do not depend on new physics.  
We will use models that incorporate event-by-event fluctuations in initial  
conditions: a `toy' wounded nucleon model and  the hadronic string  
dynamics (HSD) transport model to illustrate the effect of these  
contributions. We then argue that a study of the dependence of correlations 
on the centrality bin definition as well as the bin size may distinguish  
between these `trivial' correlations and correlations arising from  
`new physics'.
 
The paper is organized as follows. In section II  the main  
observables are introduced. In sections III and IV we study   
system size fluctuations and the resulting centrality dependence of 
correlations of two disconnected regions in momentum space within two different  
models: the Glauber Monte Carlo model (with no hadronic  
re-interactions or initial state dynamics) and the 
HSD transport model. Section V summarizes our study.

\section{Definition of Observables}  
  
The statistical properties of a particular sample  of events can  
be characterized by a set of moments or cumulants of some  
observable.  These properties depend upon a set of criterions  
which are used to select this sample. Applied to the context of  
heavy-ion collisions this translates to the construction of  
centrality bins of collision events from minimum-bias data. We  
will discuss the charged hadron multiplicities $N_A$ and $N_B$ in  
two symmetric intervals $\Delta \eta$ of pseudo-rapidity. After  
construction of the centrality bins, one can calculate the moments  
of a resulting distribution  
$P^{\eta_{gap}}_{c}(N_A,N_B;\Delta\eta)$:  
\begin{equation}  
\langle N_A^k \cdot N_B^l \rangle^{\eta_{gap}}_c ~\equiv~  
\sum_{N_A,N_B}~ N_A^k ~ N_B^l ~  
P^{\eta_{gap}}_{c}(N_A,N_B;\Delta\eta)~. \label{eq1}  
\end{equation}  
In Eq.~(\ref{eq1}) the subscript $c$ denotes a particular  
centrality bin, while the superscript $\eta_{gap}$ denotes the  
separation of two symmetric intervals $\Delta\eta$ in  
pseudo-rapidity space where particle multiplicities $N_A$ and  
$N_B$ are measured. The correlation coefficient\footnote{We use a  
different notation from Refs.  
\cite{correlation_exp1,correlation_exp2} denoting the correlation  
coefficient as $\rho$ and reserve the letter $b$ for the impact  
parameter.} is defined by  
\begin{equation}  
\rho  
~\equiv~ \frac{\langle \Delta N_A \cdot \Delta N_B \rangle^{\eta_{gap}}_c}{\sqrt  
{ \langle \left (\Delta N_A \right)^2 \rangle^{\eta_{gap}}_c ~  
\langle \left (\Delta N_B \right)^2  
\rangle^{\eta_{gap}}_c } }  
\label{eq2}  
\end{equation}  
and measures how strongly multiplicities $N_A$ and $N_B$ -- in a  
given centrality bin $c$ for pseudo-rapidity separation  
$\eta_{gap}$ -- are correlated. In Eq.~(\ref{eq2}), $\Delta N  
\equiv N-\langle N \rangle^{\eta_{gap}}_c$ and $\langle \left  
(\Delta N_A \right)^2 \rangle^{\eta_{gap}}_c~ =~\langle \left  
(\Delta N_B \right)^2 \rangle^{\eta_{gap}}_c$ for symmetric  
intervals.  
  
The recent preliminary data on forward-backward correlation  
coefficient (\ref{eq2}) of charged particles by the STAR  
Collaboration 
\cite{correlation_exp1,correlation_exp2} exhibit two striking  
features:~ a) an approximate independence on the width of the  
pseudo-rapidity gap $\eta_{gap}$~,~ b) a strong increase of  
$\rho$  with centrality.  
  
\section{Glauber Monte Carlo Model}  
  
We use the PHOBOS Glauber Monte Carlo code \cite{PGMC} coupled to  
a `toy' wounded nucleon model, referred to as GMC. The aim of this  
model is to emphasize two crucial aspects: 1)~an averaging over  
different system sizes within one centrality bin introduces   
correlations; 2)~the strength of these correlations depend on the  
criteria used for the centrality definition and on the size of the  
centrality bins.  
  
\begin{figure}[ht]  
\epsfig{file=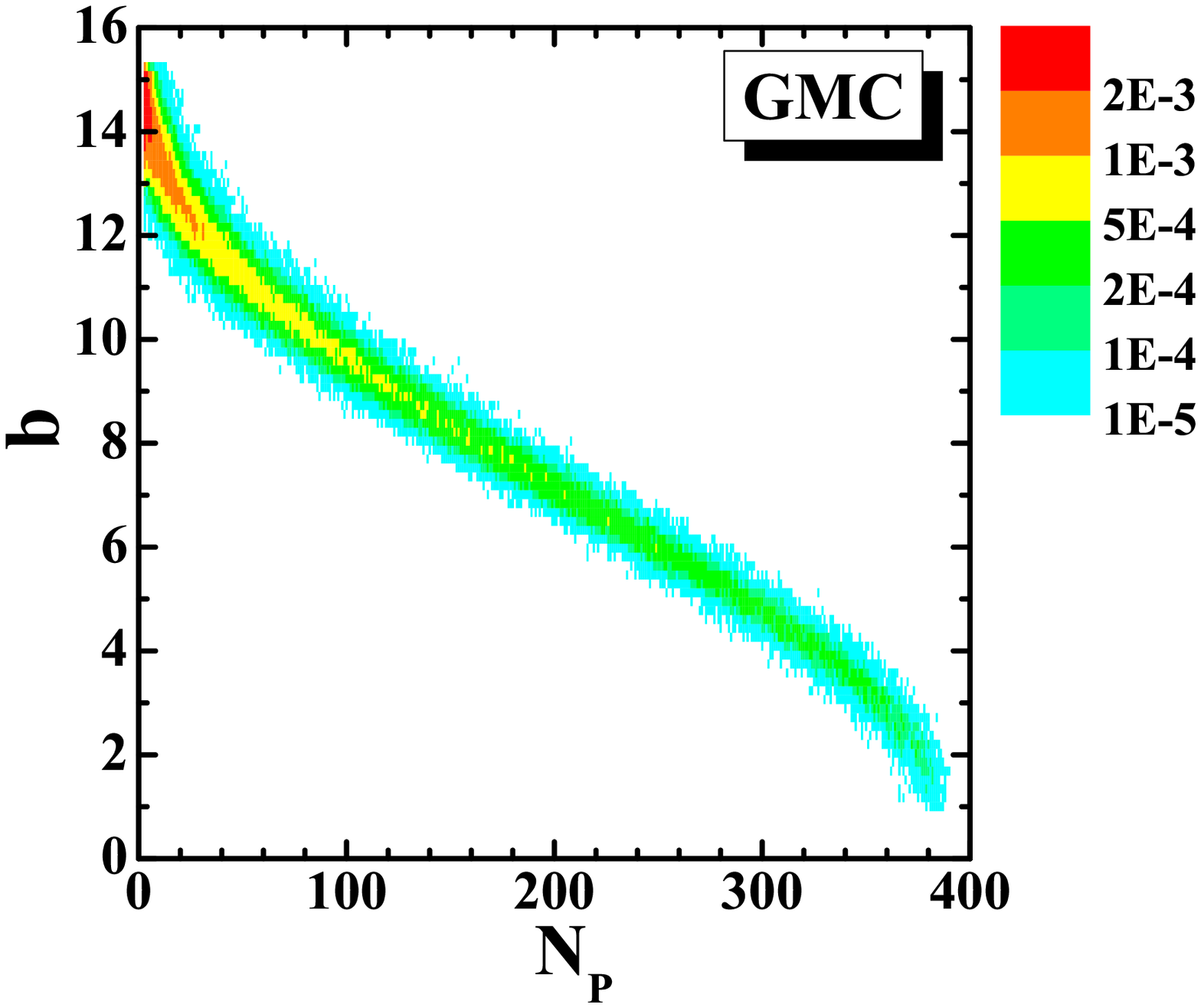,width=0.46\textwidth}  
\epsfig{file=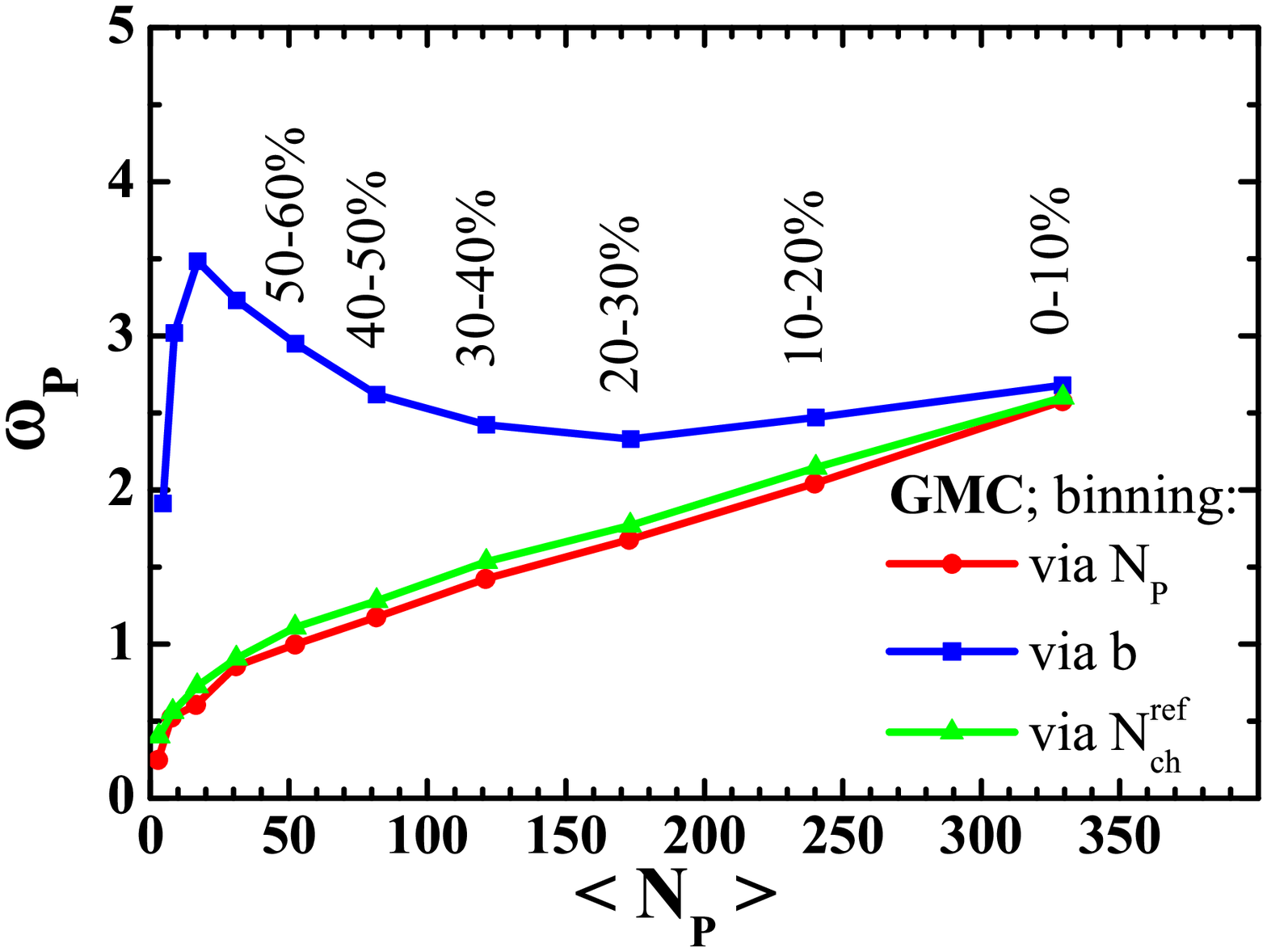,width=0.49\textwidth} \caption{(Color  
online) {\it Left}: Histogram shows the distribution of events  
with fixed number of participating nucleons $N_{P}$ and fixed  
impact parameter $b$ in Au+Au collisions at $\sqrt{s}=200$~GeV.  
{\it Right}: The scaled variance $\omega_{P}$ of the distribution  
of participating nucleons in $10 \%$ bins as defined via $b$,  
$N_{P}$, and $N_{ch}^{ref}$.} \label{GMC_fluc_Npart}  
\end{figure}

Employing the Glauber code we model the distribution of the number  
of participating nucleons, $N_{P}$, in each nucleus-nucleus  
collision for given impact parameter $b$ (cf. Fig.  
\ref{GMC_fluc_Npart}, {\it left}). This is done for Au+Au with  
standard Wood-Saxon profile and the nucleon-nucleon cross section  
of $\sigma_{NN}=42$~mb. The `event' construction proceeds then in a 
two-step process. Firstly, we randomly generate the total number of 
charged particles:  
\begin{equation}\label{GMC_nch}  
N_{ch} ~=~ \sum_{i=1}^{N_{P}} ~ n_{ch}^i~ ,  
\end{equation}  
where the number of charged particles $n_{ch}^i$ per participating  
nucleon  are generated by independently sampling a Poisson  
distribution with given mean value $\overline{n}_{ch}=10$.  
Secondly, these charged particles are randomly distributed  
according to a Gaussian in pseudo-rapidity space:  
\begin{equation}\label{GMC_dndy}  
\frac{dN_{ch}}{d\eta} ~\propto~ \exp \left( - \frac{\eta^2}{2  
\sigma_{\eta}} \right)~,  
\end{equation}  
where $\sigma_{\eta} = 3$ defines the width of the pseudo-rapidity  
distribution. Hence, in each single event there are no  
correlations between the momenta of any two particles. Note that  
numerical values of $\overline{n}_{ch}$ and $\sigma_{\eta}$ are  
fixed in a way to have a rough correspondence with the data on  
charged particle production at $\sqrt{s}=200$~GeV.  
  
In Fig. \ref{GMC_fluc_Npart} ({\it left}) we show the GMC event  
distribution in the $(b,N_{P})$-plane. For each of these events we  
randomly generate the number of charge particles $N_{ch}$ and  
their $\eta$-distribution according to Eqs.~(\ref{GMC_nch}) and  
(\ref{GMC_dndy}), respectively.  The construction of centrality  
classes can now be done in several ways. Here we focus on the  
following criterions:  via impact parameter $b$, via the number of  
participating (wounded) nucleons $N_{P}$,  and via the charged  
particle multiplicity $N_{ch}^{ref}$ in the midrapidity window  
$|\eta| <1$.  
  
In the case one chooses the number of participating nucleons $N_P$  
for centrality definition, one takes vertical cuts in  
Fig.~\ref{GMC_fluc_Npart} ({\it left}), while choosing the impact  
parameter $b$, one takes horizontal cuts.  Hence, depending on the  
centrality definition, one may assign a particular event  
(characterized by $N_{P}$ and $b$) to two different centrality  
bins.  
  
In Fig. \ref{GMC_fluc_Npart} ({\it right}) we show the resulting  
scaled variance $\omega_{P}$,  
\begin{equation}  
\omega_{P} ~\equiv~ \frac{ \langle \left( \Delta N_{P}  
 \right)^2 \rangle_c}{\langle N_{P} \rangle_c}~,  
\end{equation}  
of the underlying distribution of the number of participating  
nucleons $N_{P}$ in each centrality bin. Using the centrality  
selection via impact parameter $b$, which is only the  
theoretically available trigger,  one generally obtains a rather  
wide distribution of participating nucleons in each bin. The lines  
for centrality selections via $N_{ch}^{ref}$ and via $N_{P}$ are  
similar due to the event construction with  
Eqs.~(\ref{GMC_nch},\ref{GMC_dndy}). An interesting feature of the  
GMC model is that $\omega_{P}$ increases with centrality for the  
selection via $N_{P}$.  This conclusion of the GMC model seems to  
have a rather general origin.  
  
\begin{figure}[ht]  
\epsfig{file=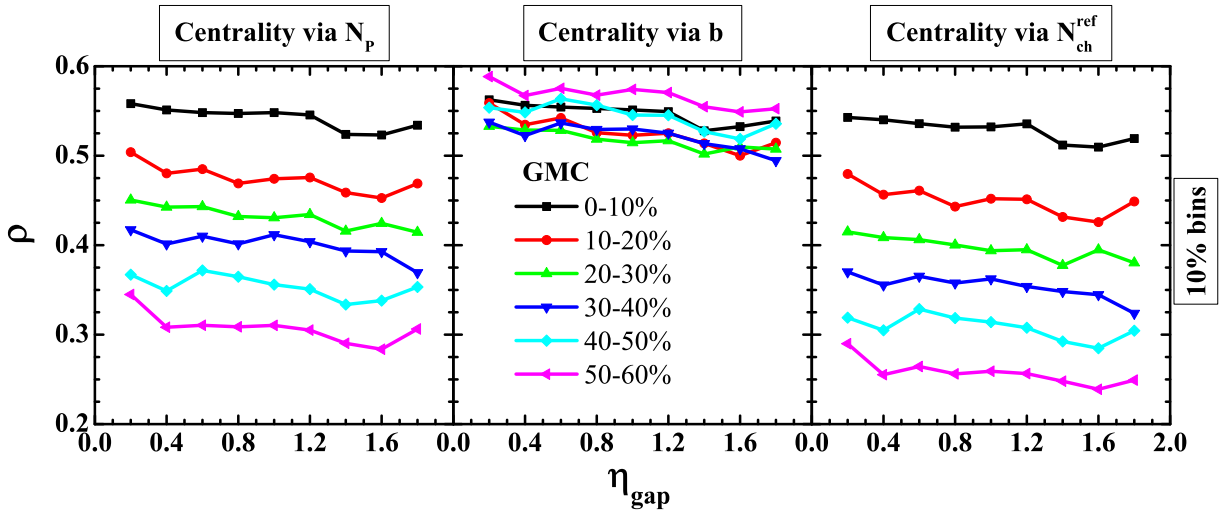,width=\textwidth} \caption{(Color online)  
The forward-backward correlation coefficient $\rho$ for $10\%$  
centrality classes defined via $N_{P}$ ({\it left}), via the  
impact parameter $b$ ({\it middle}), and via the multiplicity in  
the central rapidity region  $N^{ref}_{ch}$ ({\it right}).}  
\label{GMC_FB_corr}  
\end{figure}

We now investigate the sensitivity of the forward-backward  
correlation signal as a function of the separation $\eta_{gap}$ of  
two narrow intervals ($\Delta \eta = 0.2$) on the centrality  
definition. This is done for the 10\% centrality defined via  
$N_{P}$, via $b$, and via $N_{ch}^{ref}$. The results are shown in~Fig.~\ref{GMC_FB_corr}.  
  
In the GMC we can identify the number of participating nucleons  
$N_{P}$ with the system size, and $\omega_{P}$ as the measure for  
system size fluctuations. Having a large system as measured by  
$N_{P}$ implies a large number of charged particles $N_{ch}$. In  
GMC they are distributed independently in pseudo-rapidity space.  
Conversely, an event with small $N_{P}$ contains only few  
charged particles. By grouping the collision events into 10\%  
centrality bins one finds rather large $N_P$-fluctuations in one  
specific bin. The averaging over different states in the  
centrality bin introduces correlations between any two regions  
of pseudo-rapidity. Small systems will have few particles `on the  
left' and few particles `on the right' with respect to  
midrapidity. Large systems will have many particles `on the left'  
and many particles `on the right'.  But this just means a non-zero  
forward-backward correlations. From the definition (\ref{eq2}) one  
finds a positive correlation coefficient~$\rho$ due to averaging  
over system sizes.  
  
Note that centrality selections via $N_P$ and via $N_{ch}$ give  
essentially the same results for $\rho$ in the GMC (cf. {\it left}  
and {\it right} panels of Fig.~\ref{GMC_FB_corr}). Using the  
impact parameter $b$ for the centrality definition generates  
centrality bins with almost constant $\rho$ as seen in  
Fig.~\ref{GMC_FB_corr} ({\it middle}). This is due to a rather  
flat dependence of $\omega_{P}$ on the centrality defined via $b$,  
as shown in Fig.~\ref{GMC_fluc_Npart} ({\it right}). In the GMC  
model the apparent ordering of $\rho$ values with respect to  
centrality bins originates from the width of the underlying  
distribution in the number of wounded nucleons in each bin, i.e.  
from the values of $\omega_P$ .  
  
The measured and apparently strong forward-backward correlations  
can be accounted for by a `toy' model such as the GMC, provided it  
produces particles over the whole rapidity range and includes  
strong enough event-by-event fluctuations of $N_P$. The next  
section will show that an introduction of dynamics and hadron  
re-interactions within the HSD does not alter these conclusions  
significantly.  
  
\section{HSD Transport Model Simulations}  
  
A physically more reasonable scenario, which however also does not  
include any `new physics' (such as color glass condensate,  
quark-gluon plasma, etc.) can be obtained in the  
Hadron-String-Dynamics (HSD) transport approach  
\cite{Ehehalt,Brat97,CBRep98}.  
The HSD has been used for the description of $pA$, $\pi A$ and  
$AA$ collisions from SIS to RHIC energies \cite{exita,BratPRC04}.  
In this model, $N$, $\Delta$, N$^*$(1440), N$^*$(1535), $\Lambda$,  
$\Sigma$ and $\Sigma^*$ hyperons, $\Xi$, $\Xi^*$ and $\Omega$ as  
well as their antiparticles are included on the baryonic side,  
whereas the $0^-$ and $1^-$ octet states are incorporated in the  
mesonic sector. Inelastic baryon--baryon (and meson--baryon)  
collisions with energies above $\sqrt s_{th}\simeq 2.6$~GeV (and  
$\sqrt s_{th}\simeq 2.3$~GeV) are described by the FRITIOF string  
model \cite{FRITIOF} whereas low energy hadron--hadron collisions  
are modelled in line with experimental cross sections. As  
pre-hadronic degrees of freedom the HSD includes `effective'  
quarks (antiquarks) and diquarks (antidiquarks) which interact  
with cross sections in accordance with the constituent quark  
model.  
  
\begin{figure}[ht]  
\epsfig{file=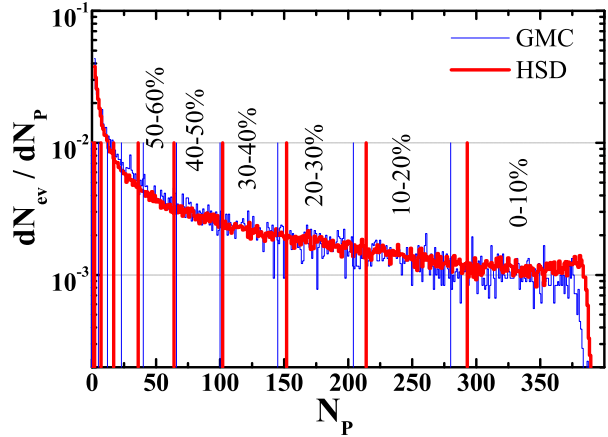,width=0.49\textwidth} \caption{(Color  
online) The HSD and GMC distributions of events over $N_P$. The  
vertical lines indicate $10\%$ centrality bins. }  
\label{distr_Npart}  
\end{figure}  
  
As before within GMC, the HSD events are generated according to a  
uniform distribution, $N_{ev}(b) \sim b$.  The resulting  
distribution of events in the $(N_{P},b)$-plane is similar to the  
GMC result depicted in Fig.  \ref{GMC_fluc_Npart} ({\it left}).  
  
In Fig. \ref{distr_Npart} we show the distribution of events with  
fixed $N_{P}$ for both models.  The vertical lines indicate $10\%$  
centrality bins as defined by the $N_{P}$ distribution. Note, that  
the peripheral part of the distribution determines also the  
centrality binning and the real bin widths. This is crucial for  
most central collisions where the number of events is small.  
Slight uncertainties in the peripheral ``tail'' of the  
distribution leads to large errors in the sizes of most central  
bins and hence to large changes in results for fluctuations and  
correlations.  
  
In contrast to the STAR data, we use in the HSD simulations the  
charged particle reference multiplicity $N_{ch}^{ref}$ in the same  
pseudo-rapidity range $|\eta|<1$ for all values of $\eta_{gap}$.  
This procedure introduces a systematic bias, since the  
pseudo-rapidity regions for the measured multiplicity in a small  
$\Delta \eta$ window (signal) and for the reference multiplicity  
partially overlap. This bias, however, is small and does not  
affect any of our conclusions.  
  
\begin{figure}[ht]  
\epsfig{file=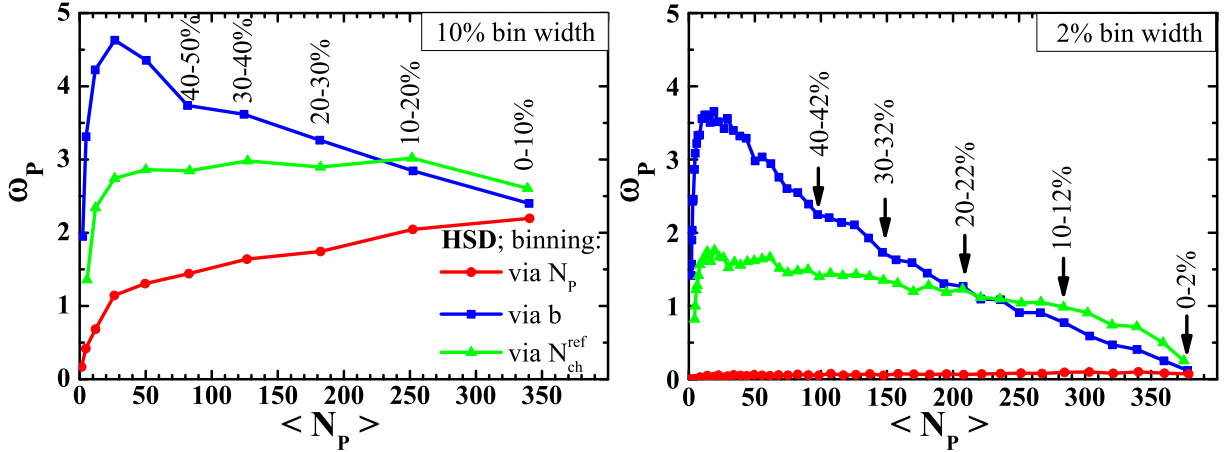,width=\textwidth} \caption{(Color online)  
The HSD results for the fluctuations $\omega_{part}$ as a function  
of the mean value $\langle N_{P}\rangle$ of the participating  
nucleons within bins as defined via $b$, $N_{P}$, and  
$N_{ch}^{ref}$.  The {\it left} panel corresponds to a 10\% and  
the {\it right} to a 2\% bin width.} \label{HSD_fluc_Npart}  
\end{figure}  
  
In Fig. \ref{HSD_fluc_Npart} we show the scaled variance of the  
underlying $N_{P}$ distribution for 10\% ({\it left}) and 2\%  
({\it right}) centrality bins defined via different centrality  
triggers within HSD. The results for 10\% bins can be compared  
with the scaled variance $\omega_P$ in the GMC model in  
Fig.~\ref{GMC_fluc_Npart} ({\it right}). Fluctuations of the  
number of participants, as well as their average values, are  
similar in both HSD and GMC models when the centrality bins are  
defined via $N_{P}$. These quantities are completely defined by  
the $N_{P}$ distribution, which is similar in both models (Fig.  
\ref{distr_Npart}).  Binning via the impact parameter $b$ in HSD,  
as well as in GMC, gives decreasing fluctuations in the  
participant number with increasing collision centrality.  The  
results for  10\% bins defined via the reference multiplicity are  
rather different in the GMC and HSD models. In GMC the charged  
multiplicity distribution is implemented according to  
Eqs.~(\ref{GMC_nch},\ref{GMC_dndy}). Hence, the results obtained  
by binning via the reference multiplicity follow the line obtained  
by binning via $N_{P}$. In contrast to the GMC, in the HSD  
simulations the average number of charged particles  
$\overline{n}_{ch}$ per participating nucleon is not a constant,  
but increases with $N_P$. Additionally, the shape of rapidity  
distribution is also different in different centrality bins. These  
two effects lead to different values of $\omega_P$ in the  
centrality bins defined via $N_{ch}^{ref}$ in the GMC and HSD  
models.  
  
One comment is appropriate here. It was argued in Ref.~\cite{BF}  
that any centrality selection in nucleus-nucleus collisions is  
equivalent to the geometrical one via impact parameter $b$. This  
result was obtained in Ref.~\cite{BF} by neglecting the  
fluctuations at a given value of $b$. Thus, different centrality  
selection criterions give indeed the same {\it average} values of  
physical observables. However, they may lead to rather different  
fluctuations of these observables in the corresponding centrality  
bins, cf. equal values of $\langle N_P\rangle$ and different  
values of $\omega_P$ for different centrality selections presented  
in Fig.~\ref{HSD_fluc_Npart}.  
  
When considering smaller centrality bins (2\% in Fig.  
\ref{HSD_fluc_Npart}, {\it right}) the fluctuations in the  
participant number become smaller and more strongly dependent on  
the definition of the binning.  
  
\begin{figure}[ht]  
\epsfig{file=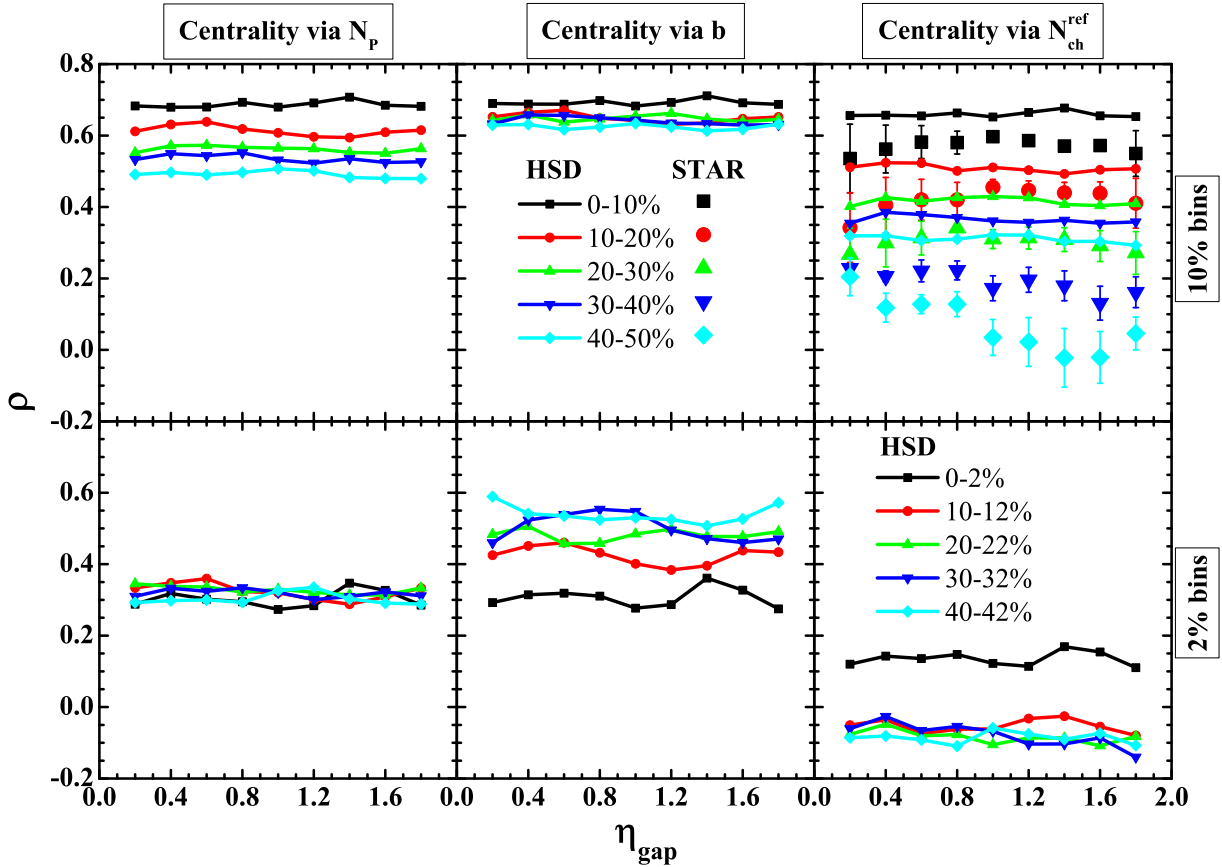,width=\textwidth} \caption{(Color online)  
The HSD results for the forward-backward correlation coefficient  
$\rho$ for 10\% ({\it top}) and 2\% ({\it bottom}) centrality  
classes defined via $N_{P}$ ({\it left}), via impact parameter $b$  
({\it center}), and via the reference multiplicity $N_{ch}^{ref}$  
({\it right}). The symbols in the {\it top right} panel present  
the STAR data in Au+Au collisions at $\sqrt{s}=200$~GeV  
\cite{correlation_exp1,correlation_exp2}. } \label{HSD_FB_corr}  
\end{figure}  
  
The Fig.~\ref{HSD_FB_corr} summarizes the dependence of  
forward-backward correlation coefficient $\rho$ as a function of  
$\eta_{gap}$ on the bin size and centrality definition within the  
HSD model. The dependence of $\rho$ on $\eta_{gap}$ is almost  
flat, reflecting a boost-invariant distribution of particles  
created by string breaking in the HSD.  
The {\it right top} panel of Fig.~\ref{HSD_FB_corr} demonstrates  
also a comparison of the HSD results with the STAR data  
\cite{correlation_exp1,correlation_exp2}. One observes that the  
HSD results exceed systematically the STAR data. However, the main  
qualitative features of the STAR data -- an approximate  
independence of the width of the pseudo-rapidity gap $\eta_{gap}$  
and a strong increase of $\rho$  with centrality -- are fully  
reproduced by the HSD simulations.  
  
\begin{figure}[ht]  
\epsfig{file=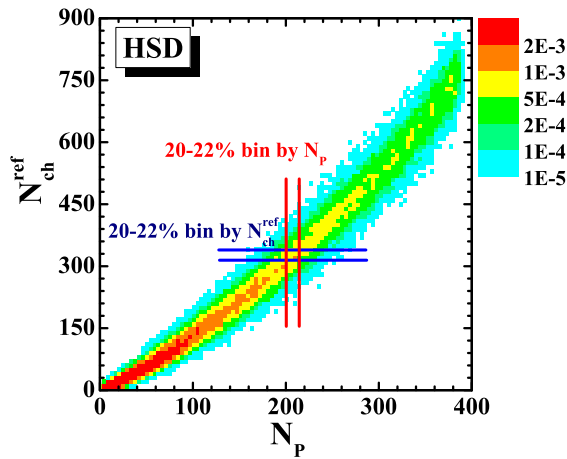,width=0.46\textwidth} \caption{(Color  
online) Histogram shows the distribution of HSD events with fixed  
number of participating nucleons $N_{P}$ and fixed reference  
charge particle multiplicity $N_{ch}^{ref}$.  The same centrality  
class (20-22\% as an example) defined in various ways contains  
different events.} \label{HSD_Npart_Nref}  
\end{figure}

The correlation coefficient $\rho$ largely follows the trend of  
the participant number  fluctuations $\omega_P$ as a function of  
centrality. The actual results, however, strongly depend on the  
way of defining the centrality bins. For instance,  choosing   
smaller centrality bins leads to weaker forward-backward  
correlations, a less pronounced centrality dependence, and a  
stronger dependence on the bin definition.  
The physical origin for this is demonstrated in  
Fig.~\ref{HSD_Npart_Nref}. As the bin size becomes comparable to  
the width of the correlation band between $N_{P}$ and  
$N_{ch}^{ref}$, the systematic deviations of different centrality  
selections  become dominant: the same centrality bins defined by  
$N_P$ and by $N_{ch}^{ref}$ contain different events and may give  
rather different values of forward-backward correlations  
coefficient $\rho$.  
  
It should be underlined that these properties are specific to the  
{\em geometric} nature of the correlations analyzed here.   If the  
observed fluctuations are of {\em dynamical} origin (for example,  
arising from the quantum fluctuations of coherent fields created  
in the first $fm/c$ of the system's lifetime as in Refs.  
\cite{correlation_cgc1,correlation_cgc2}), there are no evident  
reasons why they should strongly  depend on centrality bin  
definitions and bin sizes.  
Thus, the experimental analysis for different bin sizes and  
centrality definitions -- as performed here -- may serve as a  
diagnostic tool for an origin of the observed correlations. A  
strong specific dependence of the correlations on  bin size and  
centrality definition would signify their geometrical origin.  
  
\section{Summary}  
  
In conclusion, we have presented a study of  the system size  
event-by-event fluctuations causing the rapidity forward-backward  
correlations in relativistic heavy-ion collisions. Our analysis  
has been based on two independent models -- a `toy' wounded  
nucleon model realized as a Glauber Monte Carlo event generator  
and the microscopic HSD transport approach. We have shown that  
strong forward-backward correlations arise due to an averaging  
over many different events that belong to one 10\% centrality bin.  
In contrast to average multiplicities, the resulting fluctuations  
and correlations depend strongly on the specific centrality  
trigger. For example, the centrality selection via impact  
parameter $b$ used in most theoretical calculations and via  
$N_{ch}^{ref}$ used experimentally lead to rather different values  
of $\omega_P$ and $\rho$ and their dependence on centrality.  
  
In the HSD model the $N_P$ distribution is similar to that in the  
GMC. It includes also the fluctuations in the number of strings  
and the fluctuations in the number of hadrons from individual  
string fragmentation. The HSD simulations reveal strong  
forward-backward correlations and reproduce the main qualitative  
features of the STAR data in A+A collisions at  
RHIC energies \cite{correlation_cgc1,correlation_cgc2}.  
  
The forward-backward correlations can be studied experimentally  
for smaller size centrality bins defined by $N_{ch}^{ref}$. When  
the size of the bins decreases, the contribution of `geometrical'  
fluctuations discussed in our paper should lead to weaker  
forward-backward correlations and to a less pronounced  
centrality dependence. Let us stress that the `geometrical'  
fluctuations discussed in our paper are in fact present in all  
dynamical models of nucleus-nucleus collisions.  Thus, they should  
be carefully subtracted from the data  before any discussion of  
new physical effects.  
We hope that a future experimental analysis in the direction  
examined here will clarify wether the observed correlations by the  
STAR Collaboration at RHIC contain really additional contributions  
from `new physics'.  
  
\begin{acknowledgments}  
\noindent The authors are grateful for the useful discussions with  
W.~Broniowski, M.~Ga\'zdzicki, M.~Mitrovski, T.~Schuster, P.~Steinberg. 
G.T. acknowledges the LOEWE foundation,the 'HIC for FAIR' program and 
the ITP and  
FIAS of JW Goethe University for support. This work was in part  
supported by the Program of Fundamental Researches of the  
Department of Physics and Astronomy of NAS, Ukraine.  
\end{acknowledgments}  
  

\end{document}